\documentclass[10pt,twocolumn,superscriptaddress,floats,showpacs,nobalancelastpage,longbibliography,prb]{revtex4-2}

\usepackage{amsmath,amsfonts, amssymb, amsthm, dsfont}
\usepackage{yfonts}
\usepackage{bm}
\usepackage{mathrsfs}
\usepackage{graphicx}
\usepackage{verbatim}
\usepackage[pdfencoding=auto, psdextra]{hyperref}
\hypersetup{linktocpage}
\hypersetup{colorlinks=true,citecolor=blue,linkcolor=blue, urlcolor=blue}
\usepackage{enumitem}
\usepackage{wasysym}
\usepackage{xcolor}
\usepackage{tikz}
\usepackage{bbold}
\usepackage[caption=false]{subfig}
\usepackage{times}

\newcommand{\comments}[1]{}

\renewcommand{\cal}[1]{\mathcal{#1}}

\def\U{\mathrm{U}(1)}
\def\C{\mathcal{C}}

\def\Z{\mathbb{Z}}

\makeatletter
\def\l@subsubsection#1#2{}
\makeatother

\definecolor{OliveGreen}{cmyk}{0.64, 0, 0.95, 0.40}

\begin{document}


\title{Chiral spin liquid in a generalized Kitaev honeycomb model with $\mathbb{Z}_4$ 1-form symmetry}


\author{Yu-Xin Yang}
\affiliation{Center for Neutron Science and Technology, Guangdong Provincial Key Laboratory of Magnetoelectric Physics and Devices, School of Physics, Sun Yat-sen University, Guangzhou 510275, China}

\author{Meng Cheng}
\email{m.cheng@yale.edu}
\affiliation{Department of Physics, Yale University, New Haven, Connecticut 06511-8499, USA}

\author{Ji-Yao Chen}
\email{chenjiy3@mail.sysu.edu.cn}
\affiliation{Center for Neutron Science and Technology, Guangdong Provincial Key Laboratory of Magnetoelectric Physics and Devices, School of Physics, Sun Yat-sen University, Guangzhou 510275, China}

\date{\today}

\begin{abstract}

We explore a large $N$ generalization of the Kitaev model on the honeycomb lattice with a simple nearest-neighbor interacting Hamiltonian. In particular, we focus on the $\mathbb{Z}_4$ case with isotropic couplings, which is characterized by an exact $\mathbb{Z}_4$ one-form symmetry. Guided by symmetry considerations and an analytical study in the single chain limit, on the infinitely long cylinders, we find the model is gapped with an extremely short correlation length. Combined with the $\mathbb{Z}_4$ one-form symmetry, this suggests the model is topologically ordered. To pin down the nature of this phase, we further study the model on both finite and infinitely long strips, where we consistently find a $c=1$ conformal field theory (CFT) description, suggesting the existence of chiral edge modes described by a free boson CFT. Further evidence is found by studying the dimer correlators on infinitely long strips. We find the dimer correlation functions show a power-law decay with the exponent close to 2 on the boundary of the strip, while decay much faster in the bulk. Combined with the topological entanglement entropy extracted from cylinder geometry, we identify the spin liquid is chiral and supports a $\mathrm{U}(1)_{-8}$ chiral topological order. A unified perspective for all $\mathbb{Z}_N$ type Kitaev models is also discussed.

\end{abstract}

\maketitle

\section{Introduction} 


Topological phase is a long-range entangled phase of matter beyond the traditional Landau symmetry breaking paradigm~\cite{Wen1990,Wen2017}. Originally discovered in the electronic Quantum Hall Effect, topological phases have also found realizations in interacting spin systems, under the name of quantum spin liquid states~\cite{Savary2016,Zhou2017}.
The search for quantum spin liquid supporting gapped excitation with anyonic statistics is of paramount importance in quantum many-body physics, with potential applications in topological quantum computation~\cite{Kitaev2003}.
Traditionally, a common type of model systems to search for topological spin liquids is spin-$1/2$ Heisenberg antiferromagnets on various lattices~\cite{Yan2011,Zhu2015},
which is partially motivated by the resonating valence bond picture~\cite{Fazekas1974} and
the fact that the small spin quantum number could give rise to strong quantum fluctuations. Later on, the model search was generalized in two different directions, i.e., the so-called large spin generalization and large $N$ generalization~\cite{Auerbach1994}. 
In the former generalization, the symmetry group (SU($2$) here) remains unchanged but the local degree of freedom becomes a higher dimensional representation of the symmetry group. In the later case, the symmetry group itself is enlarged from SU($2$) to SU($N$) with $N\ge 3$.
While the large spin generalization may land in the classical limit with quantum fluctuation suppressed, an example of which is the Heisenberg models with high spin on the kagome lattice where 
spontaneous symmetry breaking ordered phases appear~\cite{Liu2015,Picot2016}, the large $N$ generalizations can have stronger quantum fluctuation, and induce various interesting quantum spin liquid states, e.g., SU($N$) chiral spin liquids on triangular and square lattices~\cite{Nataf2016,Chen2020}.

Another frontier of spin liquid physics emerges since the discovery of spin-$1/2$ Kitaev model on the honeycomb lattice~\cite{Kitaev2006}, which is an exactly solvable model with bond dependent Ising type nearest neighbor couplings. The Kitaev model contains a rich phase diagram with both gapped and gapless quantum spin liquid, partially originated from the exact $\Z_2$ 1-form symmetry, and has drawn broad attention from both experimental side and theory side~\cite{Hermanns2018,Takagi2019,Trebst2022}. The physics in Kitaev model becomes even richer when considering various other terms present in the Kitaev material~\cite{Zheng2017,Liu2018,Janssen2019,Lin2021}, which however would also spoil the exact 1-form symmetry in Kitaev model. Thus, theoretically, it would be interesting to search for generalizations of Kitaev model without breaking the conserved quantities and 1-form symmetry inherent in this model.

Indeed, the spin-$1/2$ Kitaev model can be generalized while keeping the 1-form symmetry intact. In the same spirit as the large spin generalization of Heisenberg antiferromagnets, the Kitaev model has a high spin generalization~\cite{Baskaran2008}. Via replacing the spin-$1/2$ operators with generators of SU($2$) group in higher dimensional representations, the high spin Kitaev model retains the $\Z_2$ 1-form symmetry and associated local conserved quantities. It has become clear that properties of the high spin Kitaev model would depend on whether the spin is integer or half integer~\cite{Lee2019,Lee2020,Lee2021,Dong2020} due to an anomaly with the 1-form symmetry~\cite{Ma2023,Liu2024}, and potential experimental realization of the high spin Kitaev model is under rapid progress~\cite{Stavropoulos2019,Gu2024}. However, more importantly for this work, the high spin generalizations all have the same $\Z_2$ 1-form symmetry, and thus it is perhaps not surprising to find that the spin liquid phases realizable in this class of models are in a sense conventional~\cite{Ma2023,Liu2024}.

A different way of generalizing the Kitaev model, i.e, the large $N$ generalization, has also been proposed but is much less studied~\cite{Barkeshli2015}. This generalization stems from the fact the bond dependent Ising interaction belongs to the $\Z_2$ Clifford algebra, which has a natural generalization to the $\Z_N$ case, similar to one dimension where the $\Z_2$ Ising chain can be naturally generalized to $\Z_N$ Potts chain. As one may expect, the $\Z_N$ Kitaev models have higher symmetry than the $\Z_2$ case, i.e., a $\Z_N$ 1-form symmetry, and thus more exotic spin liquids may be possible~\cite{Chen2024}.
As pointed out in Ref.~\cite{Barkeshli2015} (also discussed in later sections), one of the key ingredients in this generalization is that the $\Z_N$ 1-form symmetry guarantees topological degeneracy in each energy level, which is absent in the $\Z_2$ case.


Motivated by above reasoning, some of us (and collaborators) have taken a step in studying the $\Z_3$ Kitaev model numerically, and identified an exotic U$(1)_{12}$ chiral spin liquid in this system~\cite{Chen2024}. This was surprising given that there is only two-body nearest neighbor interaction. Here in this work, we shall further explore the $\Z_N$ Kitaev model, and focus on the $N=4$ case. As pointed out in Ref.~\cite{Barkeshli2015,Ellison2022} (also detailed in later sections), the even and odd $N$ Kitaev models have different ground state degeneracy protected by $\Z_N$ 1-form symmetry, and the $\Z_4$ case is in fact the first example of non-modular topological order (the meaning of ``non-modular'' is explained in later sections)~\cite{Ellison2022}. Thus it is highly interesting to explore the possible exotic spin liquid phase in the $\Z_4$ Kitaev model.

This manuscript is organized as follows. In Sec.~\ref{sec:2Dmodel}, we introduce the model and detail the symmetry properties, with an emphasize on comparing them with the $\Z_3$ case. 
Then in Sec.~\ref{sec:cylinder} and \ref{sec:strip}, we study the two-dimensional (2D) model using matrix product state based numerical techniques, gaining evidence for a gapped chiral spin liquid with gapless edge. We summarize the numerical results and symmetry considerations to infer the chiral topological order in Sec.~\ref{sec:topo_order}. After a brief discussion about generic $\Z_N$ Kitaev models, we conclude in Sec.~\ref{sec:ZN_case}.
In Appendix~\ref{anyontheory}, we briefly review the $\Z_N^{(p)}$ anyon theory.
In Appendix~\ref{sec:Z4Chain}, we provide a analytical understanding of the 1D limit of the $\Z_4$ Kitaev model, which could be useful for a coupled wire analysis and may be of independent interest.

\section{$\Z_4$ Kitaev Model and symmetries} 
\label{sec:2Dmodel}


Let us start by introducing the $\Z_4$ Kitaev model, which is defined on a hexagonal lattice with a 4-dimensional Hilbert space on each site. The local operators on each site form a $\Z_4$ Clifford algebra, which is generated by operators $T^x, T^y, T^z$. Specifically, in the $T^z$-diagonal basis, we have 
\begin{equation}
T^x=\sum_{a=0,1,2,3}|a+1\rangle\langle a|,\ T^z=\sum_{a=0,1,2,3} i^{a}|a\rangle\langle a|,
\end{equation}
where the addition is defined modulo 4. The $T^y$ operator is defined by the relation 
\begin{equation}
T^y = \mathrm{e}^{i\frac{\pi}{4}}{T^x}^\dag {T^z}^\dag .
\end{equation}
It is easy to verify that $({T^x})^4=({T^y})^4=({T^z})^4=1$, $T^z T^x=iT^x T^z$, generalizing the algebraic relations of Pauli matrices for spin-$1/2$ systems. Here we note that, the $T_y$ operator defined in this work differs from the $T_y$ operator defined in Ref.~\cite{Barkeshli2015} by a phase factor, the latter of which does not satisfy $(T^y)^4=1$.

\begin{figure}[h]
\centering
    \includegraphics[width=0.9\columnwidth]{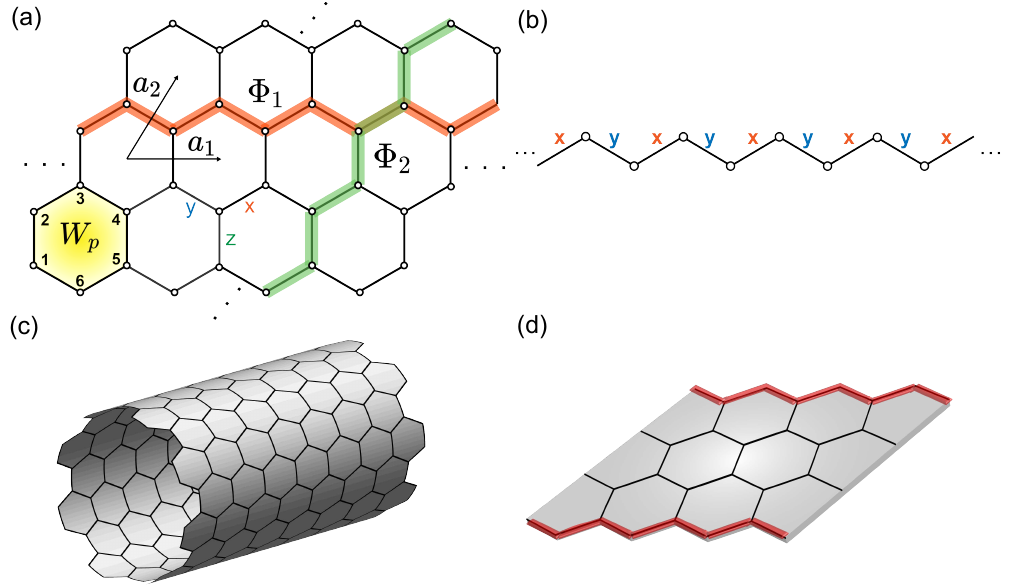}
\caption{(a) Kitaev model on the honeycomb lattice, where the conserved quantities $W_p$, $\Phi_1$, and $\Phi_2$ are highlighted. The linear size along $a_1=(1,0)^{\rm T}$, $a_2=(\frac{1}{2},\frac{\sqrt{3}}{2})^{\rm T}$ direction is denoted as $L_x$ and $L_y$, respectively, and the total system size is $2\times L_x\times L_y$.
Three different geometries are considered in this work: (b) single chain limit, (c) cylinder geometry (with cylinder axis along $a_1$ direction), (d) strip geometry. The existence of gapless edge mode identified in this work is shown in red in (d).}
\label{fig:Z4KitaevModel}
\end{figure}


We label the links of the lattice by $\alpha=x,y,z$, according to Fig.~\ref{fig:Z4KitaevModel}(a). The Hamiltonian is given by:
\begin{equation}
\label{eq:hamiltonian}
    H = \sum_{\alpha=x,y,z} J_\alpha\sum_{{\langle i,j\rangle\in \alpha-{\rm links}}} T_i^{\alpha}T_j^{\alpha}+\mathrm{h.c.}.
\end{equation}
%
%
%
%
Similar to the $\Z_3$ generalization of the original Kitaev model~\cite{Barkeshli2015, Chen2024}, the Hamiltonian of $\Z_4$ Kitaev model Eq.~\eqref{eq:hamiltonian} has complex entries in the $T^z$-diagonal basis, suggesting that time reversal symmetry is broken. This is in contrast to the orginal Kitaev model, which is time-reversal invariant, and needs three-spin interaction or external magnetic field to break time reversal symmetry.

There are a number of similarities in the $\Z_4$ Kitaev model and the $\Z_3$ Kitaev model, but some crucial differences between the two models also exist, which we now enumerate.

Firstly, associated to each hexagon, one can define a plaquette operator $W_p = T_1^x T_2^y T_3^z T_4^x T_5^y T_6^z$ with the labels of sites shown in Fig.~\ref{fig:Z4KitaevModel}(a). The $W_p$'s mutually commute and also commute with the Hamiltonian. They satisfy ${W_p}^4=1$ and can be understood as static $\Z_4$ gauge fluxes in each plaquette. Moreover, at the intersection point of three hexagons, the product of $T^x$, $T^y$, $T^z$ and its permutations are proportional to the identity. Therefore, on a closed path $\gamma$, one can define a loop operator $W(\gamma)$ by multiplying all $W_p$ operators on the plaquettes enclosed by the path $\gamma$, which would also commute with the Hamiltonian. On an open path, one can find that only the end points of the path do not commute with the Hamiltonian, suggesting that the end points carry particle-like excitations.

Secondly, one can define a Wilson loop operator along each non-contractible path on the torus. As shown in Fig.~\ref{fig:Z4KitaevModel}(a), we denote the Wilson loop operator in the $a_1$ ($a_2$) direction as $\Phi_1$ ($\Phi_2$), given by:
\begin{equation}
    \Phi_1 = \prod_{i\in A}\prod_{j\in B} T_i^z {T_j^z}^\dag,\
    \Phi_2 = \prod_{i\in A}\prod_{j\in B} T_i^y {T_j^y}^\dag,
\end{equation}
where $A$, $B$ denote the two sublattices.
For $\Z_4$ Kitaev model, the two Wilson loop operators both commute with the Hamiltonian, and satisfy $\Phi_1 \Phi_2 = -\Phi_2\Phi_1$. For the $\Z_3$ model, one instead finds $\Phi_1\Phi_2=e^{2\pi i/3}\Phi_2\Phi_1$, while for $\Z_2$ Kitaev model $\Phi_1\Phi_2=\Phi_2\Phi_1$. The non-commutativity of Wilson loop operators $\Phi_1$ and $\Phi_2$ is a general property of $\Z_N$ Kitaev model with $N\ge 3$, which is distinct from the original spin-$1/2$ Kitaev model and its higher spin generalizations. This property guarantees that every energy level has a certain degeneracy. 
Moreover, for any odd $N>1$, the degeneracy from $\Phi_1,\Phi_2$ is at least $N$, while for even $N$, this degeneracy is only $N/2$. This is one of the most important differences between $\Z_4$ and $\Z_3$ Kitaev model.

Similar to the $\Z_2$ and $\Z_3$ case, the plaquette operators $W_p$ and Wilson loop operators $\Phi_1$ and $\Phi_2$ form a so-called $\Z_4$ 1-form symmetry. 
 If the ground state is gapped, then open string operators applied to the ground state create anyonic excitations, and the 1-form symmetry determines the fusion and braiding statistics of anyons. We will thus use ``1-form symmetry group" and ``anyon theory" interchangably. As shown in Ref.~\cite{Ellison2022}, the anyon theory determined by $\Z_4$ 1-form symmetry is the $\Z_4^{(-1)}$ theory. As reviewed in Appendix \ref{anyontheory}, there are four types of anyons labeled by $[a]$ where $a=0,1,2,3$. Their exchange statistics is $\theta(a)=i^{a^2}$.
 What distinguishes the $\Z_4$ case from the previously studied $\Z_2$ and $\Z_3$ case is that the $\Z_4^{(-1)}$ theory contains a \emph{transparent boson} $[2]$. Here transparent means that the braiding of $[2]$ with any other anyon is trivial. An anyon theory with nontrivial transparent anyons is said to be ``non-modular", and can not be realized as the full topological order of a gapped ground state.   Thus there are two possibilities in a gapped state: either $[2]$ is a topologically nontrivial boson, which means in addition to $\Z_4^{(-1)}$ there must be other anyons in the system to braid nontrivially with $[2]$, or $[2]$ actually represents a local excitation, and once identifying $[2]\sim [0]$ the anyon theory is the same as a chiral semion $\Z_2^{(-\frac12)}$. Which of the two is realized will need to be answered numerically.

Moreover, there is an interesting interplay between these conserved quantities and translation symmetry:
\begin{equation}
\label{eq:relation}
    \Phi_{2,x}^\dag\Phi_{2,x+1} = \prod_y W_{p,y},
\end{equation}
where $\Phi_{2,x}$ is supported along column $x$ and the product runs over all the hexagons between column $x$ and $x+1$. In the $\Z_3$ case studied in Ref.~\cite{Chen2024}, it was found that the ground state has nontrivial flux, which renders translation symmetry being spontaneously broken for certain system size. It would be interesting to see whether similar scenario appear in the $\Z_4$ case.

Apart from the $\Z_4$ 1-form symmetry, one can get further analytical understanding of this model by studying the single chain limit, which correspond to a $\Z_4$ Kitaev chain, schematically shown in Fig.~\ref{fig:Z4KitaevModel}(b).
In the Appendix~\ref{sec:Z4Chain}, we show analytically that the $\Z_4$ Kitaev chain is dual to a $\Z_4$ clock model with extensive number of conserved quantities~\footnote{A similar observation was made in Ref.~\cite{Chen2024}, and also recently realized in the spin-1/2 case~\cite{Pujari2024}.}. 
The dual $\Z_4$ clock chain has a crticial point at $J_x=J_y$ (with $J_x,J_y\in \mathbb{R}$), which is described by a Ising$^2$ conformal field theory (CFT) with central charge $c=1$. This implies that the original $\Z_4$ Kitaev chain is also described by the same CFT at this parameter. One can then take a coupled wire analysis to infer the two dimensional phase of the $\Z_4$ Kitaev model.
%
%
This is indeed possible (as indicated by the non-local mapping in Appendix~\ref{sec:Z4Chain}), and has been worked out for the $\Z_3$ case in Ref.~\cite{Barkeshli2015}. However, since coupled wire construction typically would lead to complicated interactions between the chains, here we will not go further along this line, but only draw an intuition that the two-dimensional phase could have a chiral gapless edge mode with suitable couplings between the chains. This is especially true when we take $J_x=J_y$, where the single chain would be described by a CFT.
In the following, we will turn to the two dimensional model and study it numerically.

\section{Results on cylinders}
\label{sec:cylinder}

The $\Z_4$ 1-form symmetry guarantees that in a gapped state $W(\gamma)$ creates nontrivial, deconfined anyon excitations when $\gamma$ is an open path. However, $W^2(\gamma)$ can only create a transparent boson, which may or may not be nontrivial. Correspondingly, the algebra of non-contractible Wilson loop operators guarantees a two-fold topological degeneracy. More formally, the anomaly in the $\Z_4^{(-1)}$ symmetry is compatible with spontaneous breaking to $\Z_2^{(0)}$ or nothing in a gapped ground state.

To fully understand the topological order, one needs to first check whether the model is gapped or not in the two-dimensional thermodynamic limit. Secondly, if the system is gapped it is important to understand how the $\Z_4$ 1-form symmetry spontaneously breaks. This can be revealed by studying the ground state degeneracy on a cylinder.

Motivated by these questions, we now turn to numerical study of this model using matrix product state based infinite density matrix renormalization group (iDMRG) method~\cite{McCulloch2008}. Based on the single chain analysis, in the following we will focus on the $J_x=J_y$ case and further take the parameter $J_x=J_y=J_z=-1$, i.e., the ferromagnetic isotropic point, while leaving exploration of the complete phase diagram to future works. 

To study the bulk properties, we put the system on infinitely long cylinders, i.e., periodic boundary condition along $a_2$-direction, and study the ground state properties with varying cylinder width $L_y$. See Fig.~\ref{fig:Z4KitaevModel}(c) for illustration. We denote the cylinder with width $L_y$ as YC$L_y$.
If a finite width cylinder is viewed as a quasi-one-dimensional system, the conserved quantities $W_p$ and $\Phi_2$ are local operators while the Wilson loop operator $\Phi_1$ is not. Therefore, it is reasonable to require the variational ground state to be an common eigenstate of all $W_p$ and $\Phi_2$ operators. 

\begin{figure}[b]
    \centering
    \subfloat{\includegraphics[width=0.49\columnwidth]{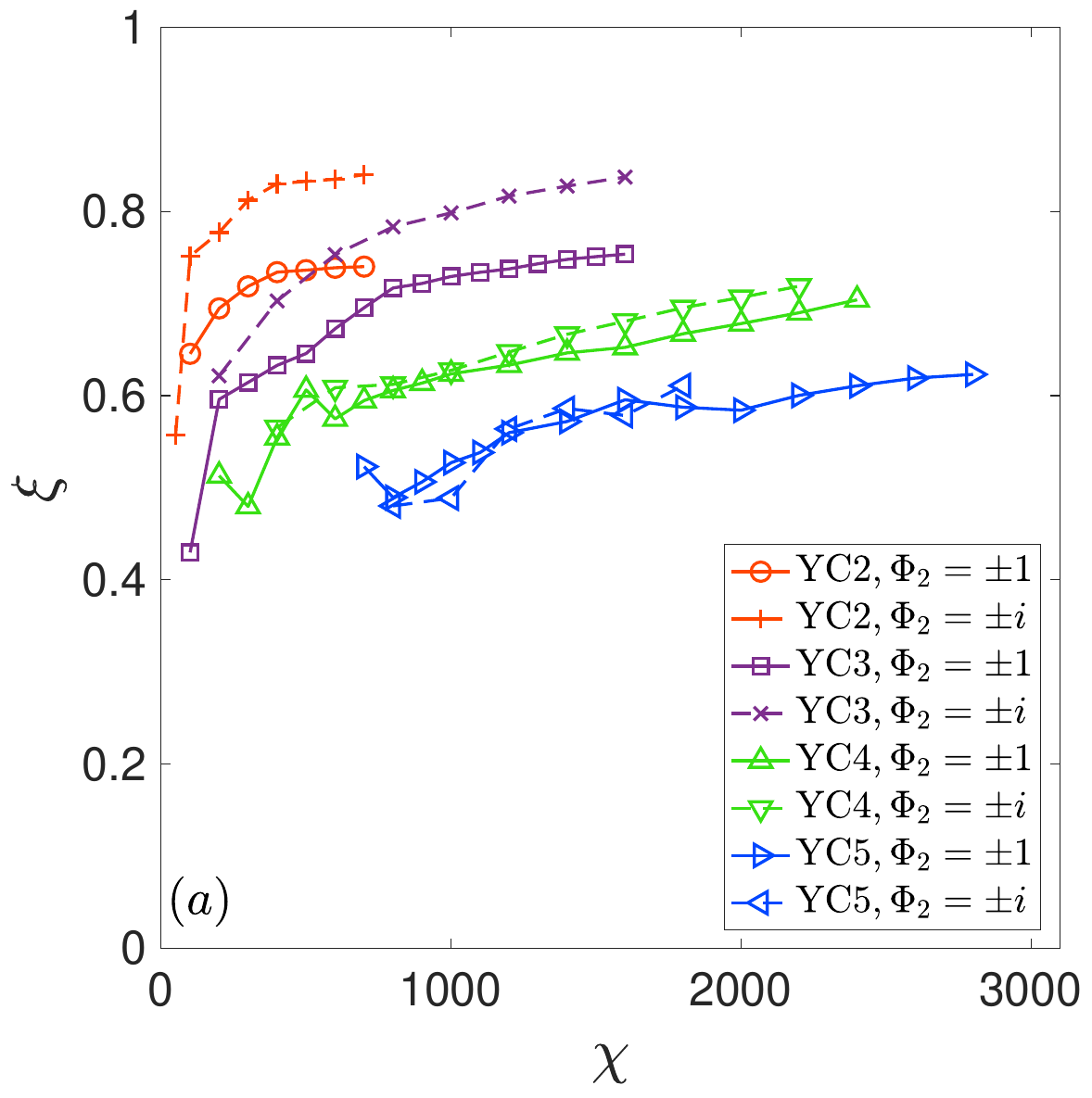}}
    \subfloat{\includegraphics[width=0.48\columnwidth]{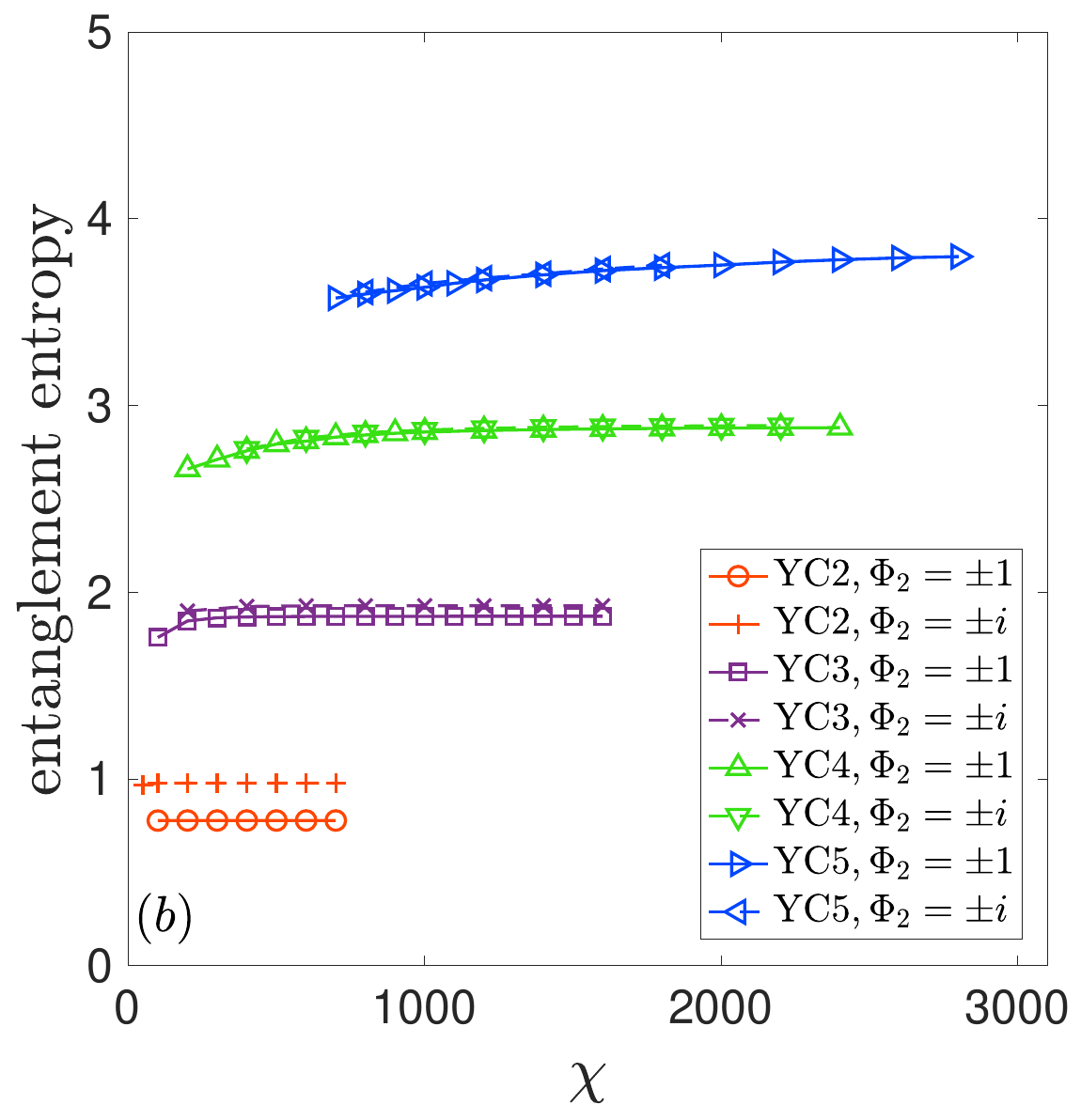}}\\
    \subfloat{\includegraphics[width=0.48\columnwidth]{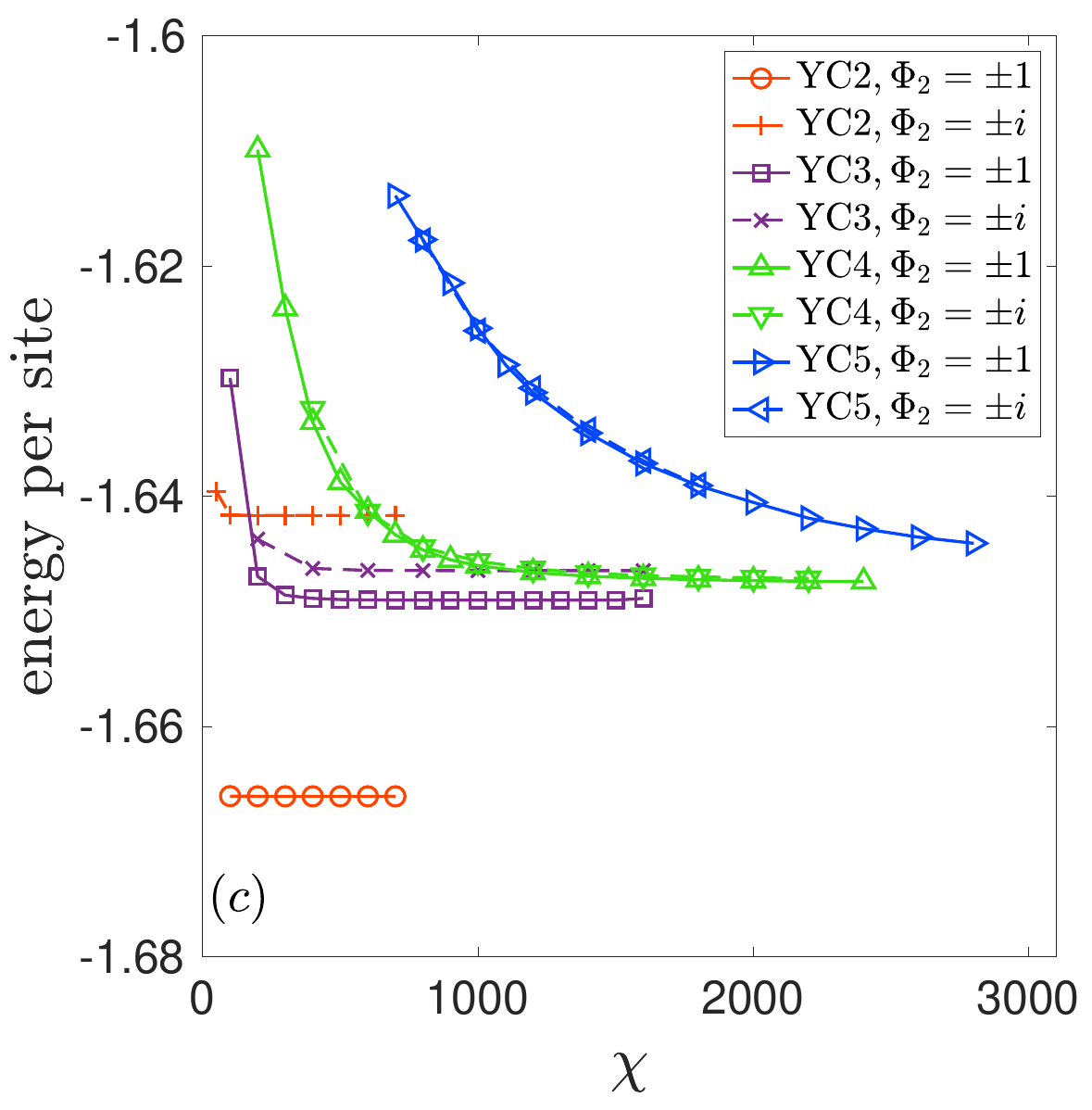}}
    \subfloat{\includegraphics[width=0.47\columnwidth]{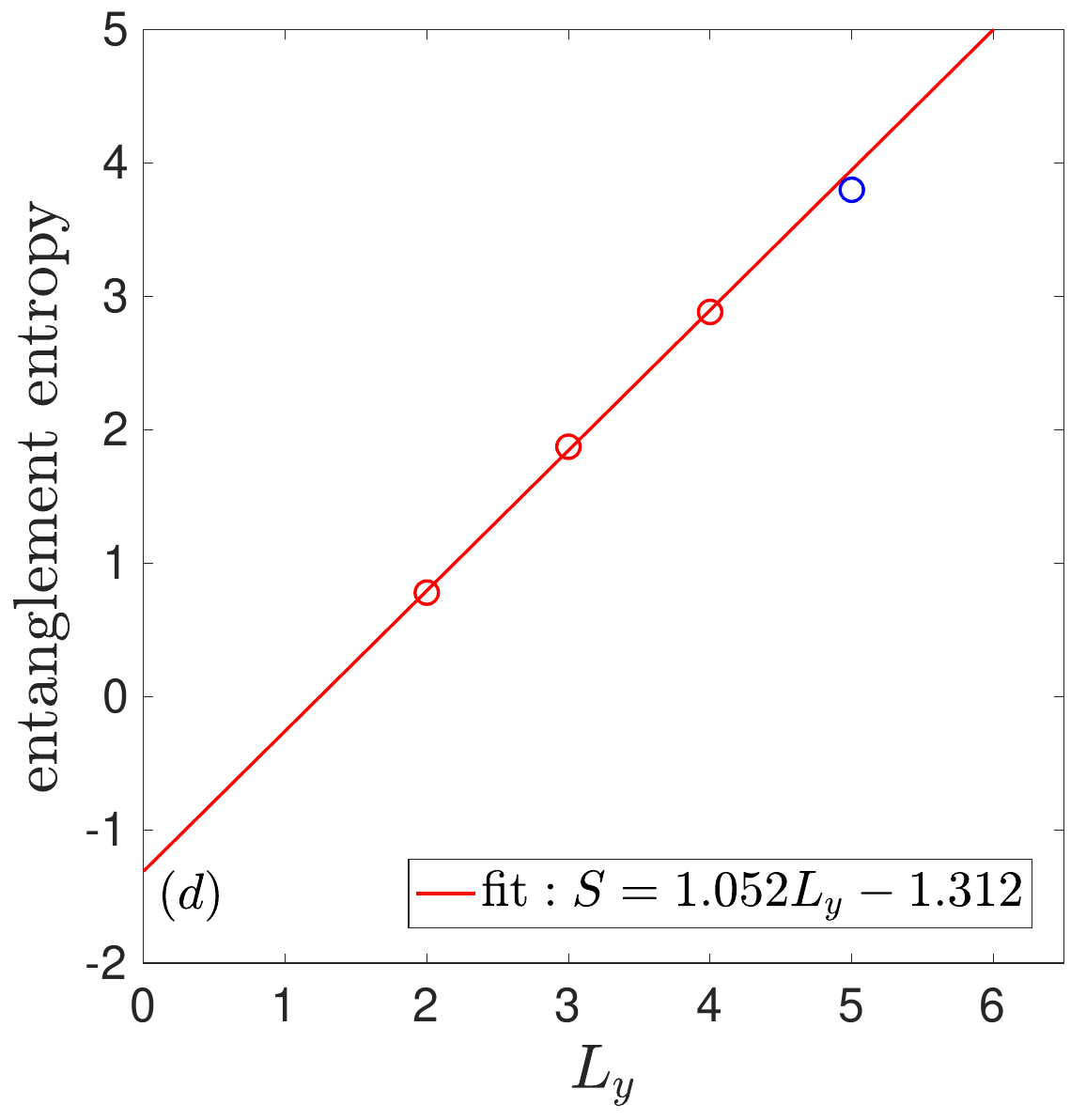}}
\caption{Numerical results on finite width cylinders. (a), (b) and (c) show the ground state correlation length, entanglement entropy, and energy density versus MPS bond dimension $\chi$ for various cylinder width $L_y$, respectively. In (c), the energy is measured with respect to the Hamiltonian Eq.~\eqref{eq:hamiltonian} without $J_{\Phi_2}$ term. In (d) we fit the entanglement entropy in $\langle \Phi_2 \rangle=\pm 1$ sector versus cylinder width $L_y$ to obtain the topological entanglement entropy, with the fitted value being $\gamma=1.312\pm 0.95$. Each data in (d) corresponds to the available data with largest bond dimension. Since the bond dimension to fully converge the $L_y=5$ result is beyond the computational capability, we do not include this data in the fitting.}
\label{fig:cylinder}
\end{figure}

Due to the anti-commutation relation between Wilson loop operators $\Phi_1$ and $\Phi_2$: $\Phi_1\Phi_2=-\Phi_2\Phi_1$, the variational ground states can be grouped into two sets, i.e., the state with $\langle \Phi_2\rangle =\pm 1$ and $\langle \Phi_2 \rangle =\pm i$. {\it A priori}, it is not clear which sector contains the true ground state.
In order to find a variational ground state with definite $\langle\Phi_2\rangle$, we add a term $\sum_x J_{\Phi_2}\Phi_{2,x}+{\rm h.c.}$ to the Hamiltonian Eq.~\eqref{eq:hamiltonian}, and choose $J_{\Phi_2}=-0.5$ $(0.5i)$ to ensure an eigenstate of $\Phi_2$ with the eigenvalue being $\langle \Phi_2\rangle = 1$ $(i)$. Then the other eigenstates of $\Phi_2$ can be obtained by inserting a $\Phi_1$ string operator.
With this trick, we have computed ground states for various $L_y$ and bond dimension $\chi$ in all $\Phi_2$ sectors, and the results are shown in Fig.~\ref{fig:cylinder}.

It turns out that, for all the cylinder widths we have considered ($L_y=2,3,4,5$), the ground states in the four $\Phi_2$ sectors all have $\langle W_p \rangle=1$. This is obtained from iDMRG simulation with MPS unit cell of both two columns and four columns, and is further confirmed by exact diagonalization on a small torus (with size $L_x=2$, $L_y=2$) and density matrix renormalization group simulation on finite size cylinders (with size $L_x=16$, $L_y=2$ and $L_x=12$, $L_y=3$). (The results presented in this section are mainly obtained with MPS unit cell of two columns.) As a consequence, from the relation between $\Phi_2$ and $W_p$ operators Eq.~\eqref{eq:relation}, one can see that $\Phi_{2,x}$ takes the same value for all $x$, and there is no spontaneous translation symmetry breaking involved. This behavior is similar to the original spin-$1/2$ Kitaev model where the ground state is also flux-free, while different from the $\Z_3$ Kitaev model where the ground state has non-zero flux. The absence of flux also indicates that in principle one can use MPS with single column unit cell to approximate the ground state. 

Due to the relatively large local spin dimension, we have restricted the cylinder width to $L_y=5$ to obtain relatively converged results. For $L_y=2,3,4$ cylinder, the results are relatively easy to converge, and with bond dimension $\chi=700\ (1600,\ 2400)$, we have achieved a truncation error $1.5\times 10^{-12},\ (1.9\times 10^{-8},\ 8.5\times 10^{-6})$. For $L_y=5$, we have pushed to bond dimension $\chi=2800$, while the truncation error is around $2.8\times 10^{-4}$.

As shown in Fig.~\ref{fig:cylinder}(a), in all $\Phi_2$ sectors, the ground state correlation lengths $\xi$ saturate with increasing bond dimension $\chi$, reaching a value smaller than 1 lattice spacing. Moreover, comparing the values for different cylinder width $L_y$, the correlation length $\xi$ even decreases with increasing $L_y$, suggesting the correlation length in the two-dimensional limit is finite. Similar to the correlation length, the entanglement entropy (measured along the entanglement cut which bipartitions the cylinder) in all four sectors also show a rapidly converging behavior with increasing $\chi$, as shown in Fig.~\ref{fig:cylinder}(b).

A salient feature one can observe from Fig.~\ref{fig:cylinder}(a) and (b) is that the correlation length and entanglement entropy in the $\langle \Phi_2\rangle =\pm 1$ and $\langle \Phi_2\rangle =\pm i$ sectors are close to each other and become more degenerate with increasing cylinder width $L_y$. This suggests that ground states in the four $\Phi_2$ sectors are actually topologically degenerate. To further support this intuition, in Fig.~\ref{fig:cylinder}(c) we plot the variational energy (per site) versus MPS bond dimension $\chi$ for all cylinder widths. One can observe that, although the energy density in $\langle\Phi_2\rangle=\pm 1$ is slightly lower than that of the other sectors, with increasing $L_y$, the energy difference becomes increasingly small. Thus we find that although the $\Z_4$ 1-form symmetry only protects two exactly degenerate ground states, the four $\Phi_2$ sectors are in fact degenerate in the two-dimensional limit. 
In other words,  the entire $\Z_4^{(-1)}$ anyon theory emerges in the system. The braiding non-degeneracy of 2D topological order then requires the existence of more anyons to be a physical theory.
 

Having established the gapped nature of the ground state, we now move on to characterize the topological order. One quantity we can immediately extract is the topological entanglement entropy (TEE), which is a subleading term to the area law of entanglement entropy $S=\alpha L_y - \gamma$. Depending on the topological sector, it is known that $\gamma = \ln\mathcal{D}/d_a$, where $\mathcal{D}$ is the total quantum dimension of the topological order and $d_a$ is the quantum dimension of anyon $a$ threading through the cylinder. The result for ground state in the identity sector is shown in Fig.~\ref{fig:cylinder}(d), from which we obtain the total quantum dimension to be $\mathcal{D}={\rm exp}(1.312)\approx 3.71$.

Here we note that, as shown in Fig.~\ref{fig:cylinder}(b), the entanglement entropy for the ground state in all the $\Phi_2$ sectors become closer to each other with increasing $L_y$, suggesting that the associated anyon quantum dimension is $1$, as expected from the $\Z_4$ 1-form symmetry. On the other hand, both braiding non-degeneracy and the numerical value of the total quantum dimension suggests that dimension of ground state manifold is more than 4, indicating that in each $\Phi_2$ sector there are additional topological degenerate ground states. Indeed, on a width $L_y=3$ cylinder, with slightly large bond dimension we have found a state with energy, entanglement entropy and correlation length all being close to that of the ground state (with $\langle \Phi_2 \rangle=\pm 1$), while being orthogonal to the ground state (as measured by the overlap from transfer matrix).

The single chain analysis in Appendix~\ref{sec:Z4Chain} suggests that the topological phase may have edge states, which, based on the Li-Haldane conjecture, would lead to linear dispersing mode in the low-energy part of the entanglement spectrum~\cite{Li2008}. However, similar to the $\Z_3$ Kitaev model, the conserved quantities $W_p$ lead to extensive degeneracy in the entanglement spectrum, making it hard to find useful information for possible edge mode (data not shown). Therefore, to gain more information for this phase, in the following we will move on the strip geometry (depicted in Fig.~\ref{fig:Z4KitaevModel}(d)), where one should be able to find signatures for edge modes on the physical edge, if exist.

\section{Results on strips}
\label{sec:strip}

\begin{figure}[ht]
\centering
    \subfloat{\includegraphics[width=0.48\columnwidth]{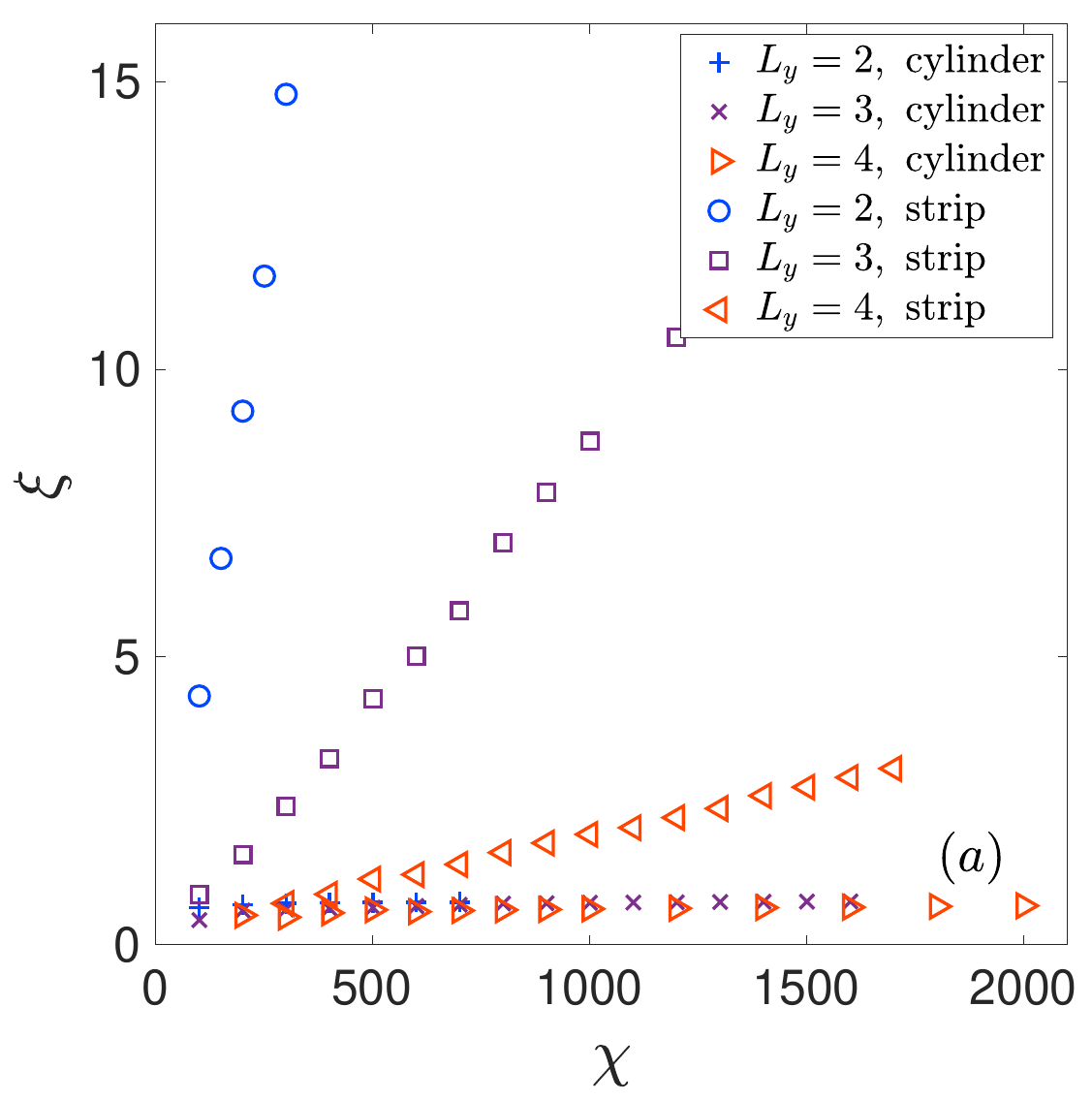}}
    \subfloat{\includegraphics[width=0.495\columnwidth]{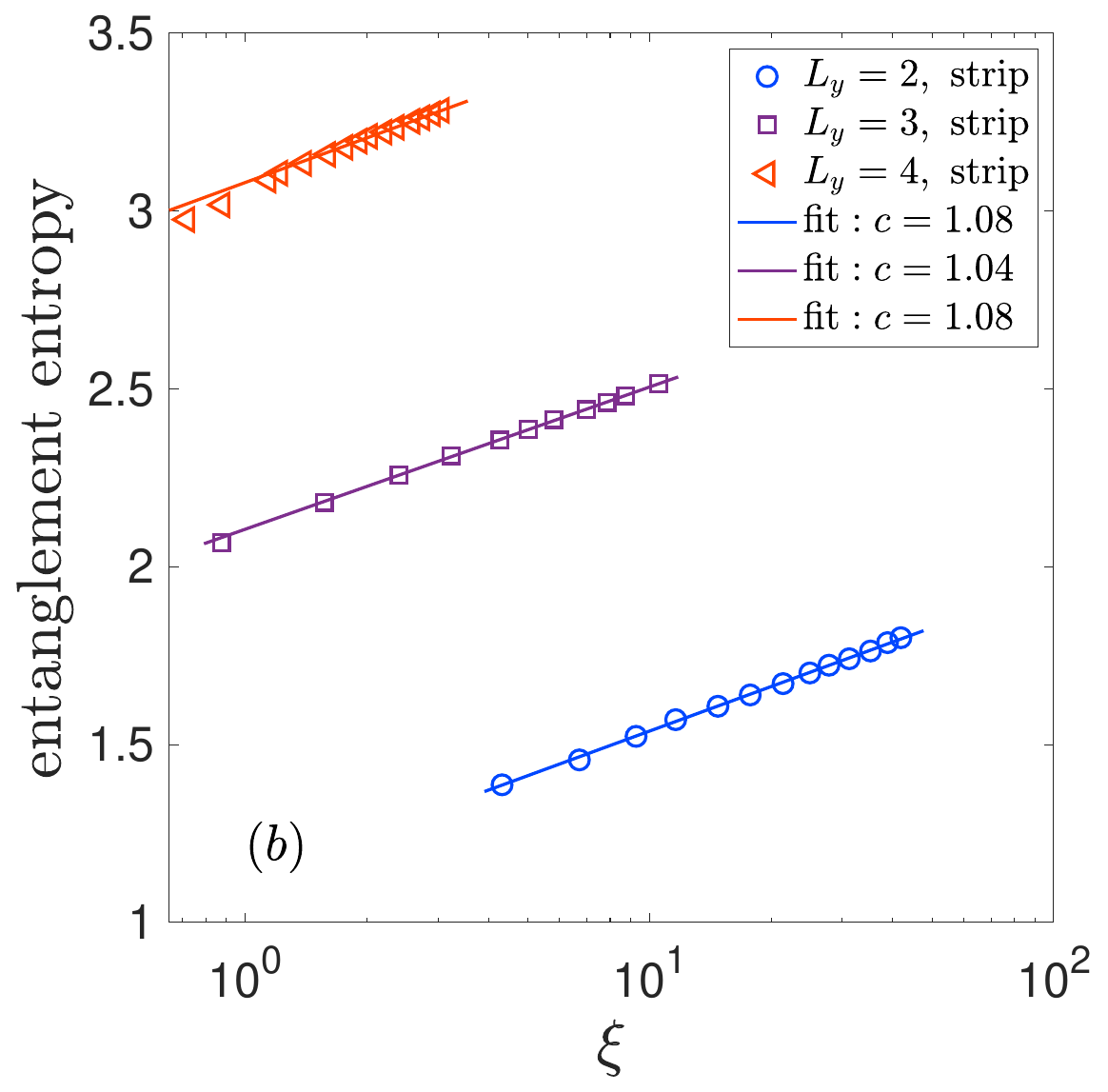}}
\caption{(a) Comparison of ground state correlation length for cylinder and strip geometries in the $\Phi_2=1$ sector. While $\xi$ saturates with increasing $\chi$ on cylinders, it diverges with $\chi$ on strips for all the widths. (b) Using finite entanglement scaling on the strip geometry, we consistently obtain a central charge $c=1$ on all the strips.}
\label{fig:infinite_strip}
\end{figure}

On a strip geometry, if there are gapless edge modes, then in the low energy regime, the system would behave as a one-dimensional gapless system, which is often described by a (1+1)D CFT. A wealth of numerical approaches for detecting (1+1)D CFT have been developed using tensor networks, which we will employ in the strip geometry.

Here, for the $\Z_4$ Kitaev model on a strip (see Fig.~\ref{fig:Z4KitaevModel}(d)), the ground state manifold would have extensive degeneracy, following the same reasoning as the $\Z_3$ case~\footnote{In fact, every energy level is degenerate, and the degeneracy is exponential in length of the strip $L_x$. See Ref.~\cite{Chen2024} for more details.}. In order to avoid an Schr{\"o}dinger cat state, we have added a $\sum_x J_{\Phi_2}\Phi_{2,x}+{\rm h.c.}$ to the Hamiltonian to break the degeneracy.

In Fig.~\ref{fig:infinite_strip}(a), we compare the ground state correlation lengths on the infinitely long strip and cylinder geometry. In both cases, the ground states satisfy $\langle \Phi_2 \rangle=1$ and $\langle W_p \rangle=1$. One can clearly see that the correlation lengths for the width $L_y=2,3$ strips are significantly larger than that on the cylinder geometry, and show a diverging behavior with increasing bond dimension $\chi$. On the $L_y=4$ strip, due to the entanglement area-law limitation with larger width, the correlation length grows less quickly, which nevertheless still increases almost linearly with $\chi$. Taken together, the data suggests that for $L_y=2,3,4$ strips, the system is indeed gapless, in contrast to the cylinder geometry.

To characterize the gapless mode, one informative quantity is the central charge $c$, which can be conveniently obtained from the finite entanglement scaling analysis~\cite{Pollmann2009}. In Fig.~\ref{fig:infinite_strip}(b), we fit the entanglement entropy versus the correlation length using $S=\frac{c}{6}\mathrm{ln}\xi+\mathrm{const}.$, where the entanglement cut bipartition the strip into left and right half. From the fit, we obtain a central charge $c\approx 1$ for all strips we have considered. Here we note that, for the $L_y=4$ case, we have used the data with a large $\chi$ to do the fitting, since the larger width would require larger bond dimension to enter the scaling regime.

\begin{figure}[hb]
\centering
\includegraphics[width=0.75\columnwidth]{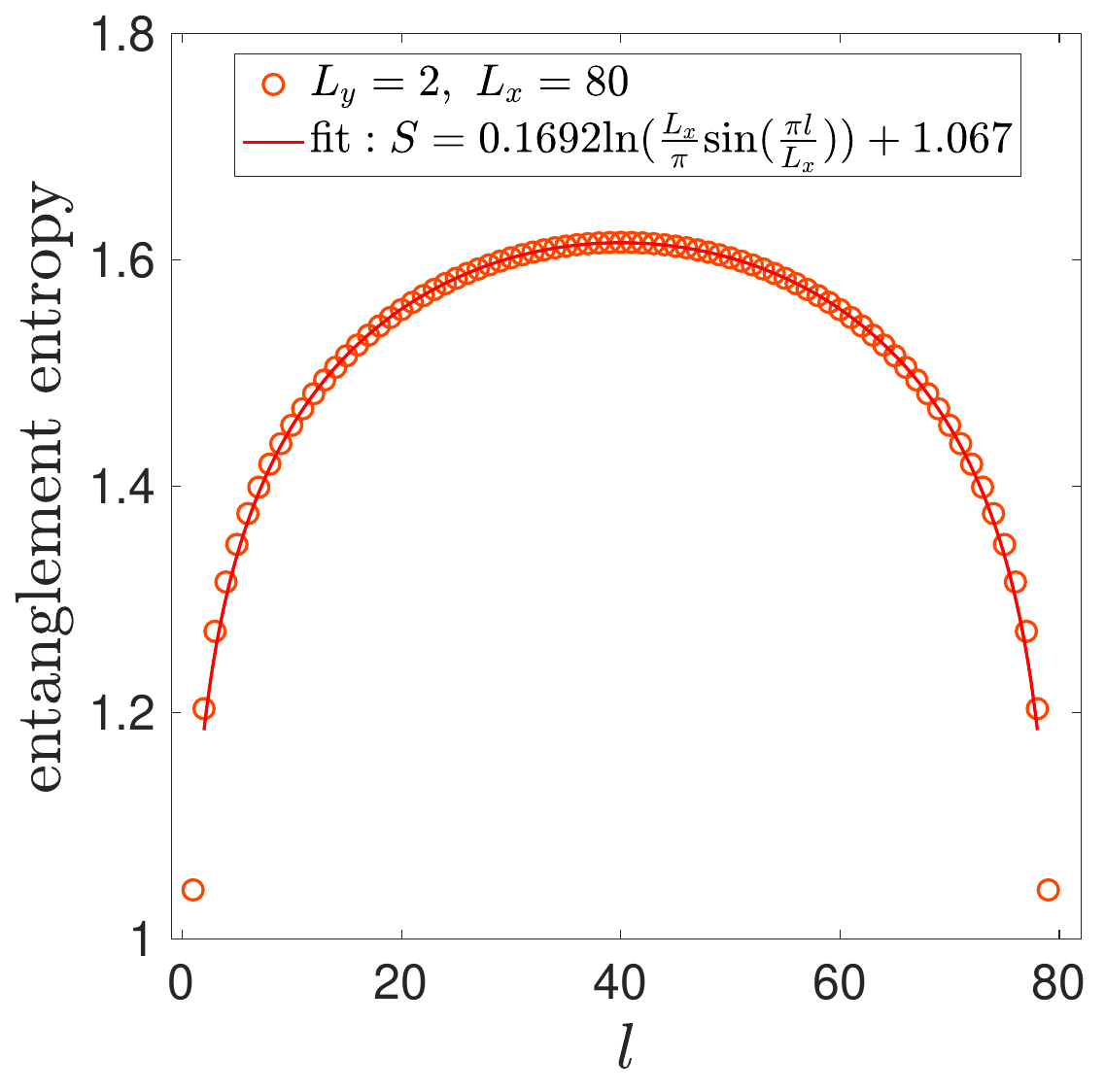}
\caption{Entanglement entropy on a $L_x=80, L_y=2$ strip with open boundary condition along the $a_1$ and $a_2$ direction. The bond dimension is $\chi=1200$. Using the Calabrese-Cardy formula, we fit the entanglement entropy versus subsystem linear size $l$ (in $a_1$ direction) and obtain a central charge close to $c=1$.}
\label{fig:finite_strip}
\end{figure}

Before moving further, we note that similar scaling of entanglement entropy versus correlation length was also obtained on the $L_y=2$ strip in the $\langle \Phi_2 \rangle =i$ sector, with the same central charge $c=1$ (data not shown). Moreover, we have also observed similar $c=1$ central charge on a finite long $L_y=2$ strip, as shown in Fig.~\ref{fig:finite_strip}. Here we remark that the central charge on the finite strip requires a relatively long strip to converge the result, and a short strip would lead to suspiciously smaller central charge (e.g., a fitted $c=0.96$ for $L_x=48,\ L_y=2$ strip), which we ascribe to a finite size effect.

\begin{figure}[h]
\centering
    \subfloat{\includegraphics[width=0.48\columnwidth]{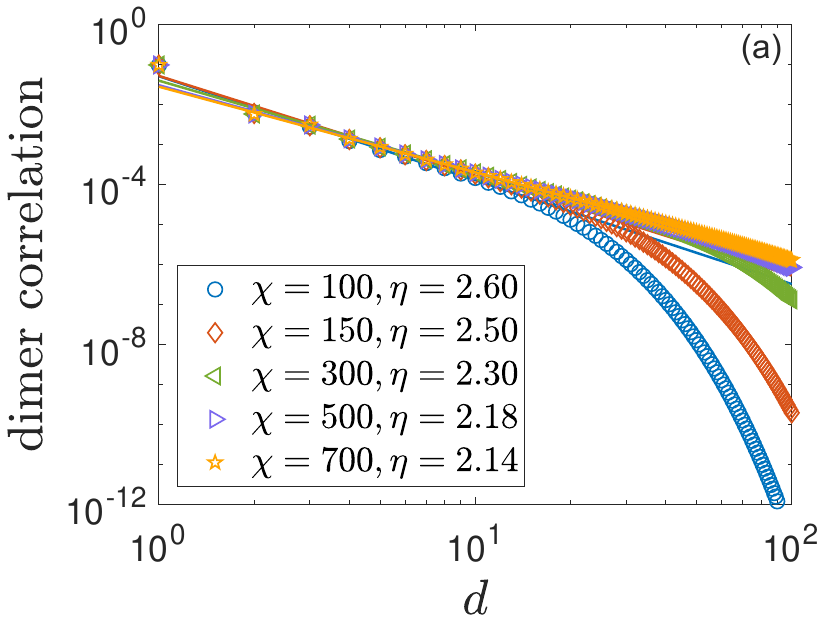}}
    \subfloat{\includegraphics[width=0.49\columnwidth]{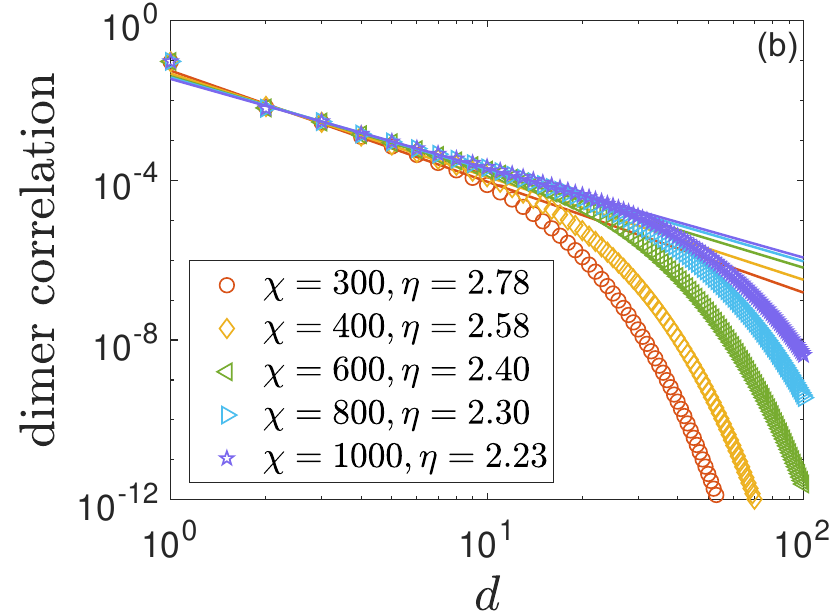}}\\
    \subfloat{\includegraphics[width=0.45\columnwidth]{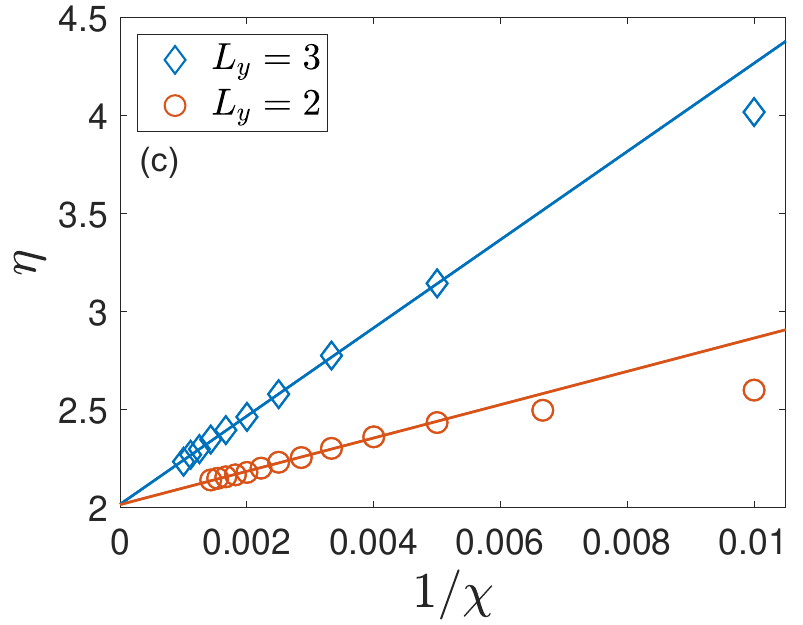}}
    \subfloat{\includegraphics[width=0.48\columnwidth]{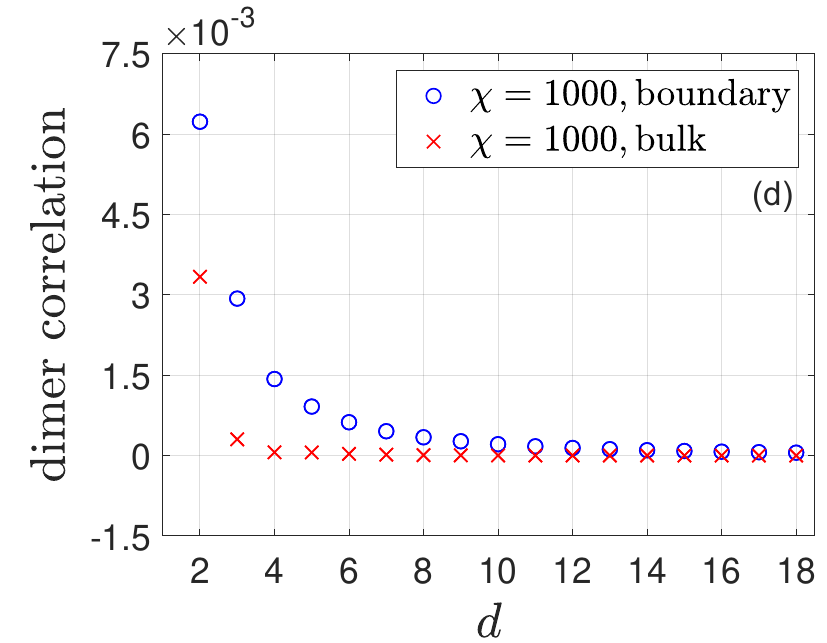}}
\caption{Dimer correlation on the strip geometry. (a) On the $L_y=2$ strip, dimer correlation decays algebraically in a range limited by the maximal correlation length induced by finite bond dimension $\chi$. (b) Similar algebraically decaying dimer correlation was also observed on the boundary of $L_y=3$ strip. (c) Extrapolating the critical exponent of dimer correlation to the infinite $\chi$ limit, we obtain an exponent close to $2$.  (d) In the short distance regime of a $L_y=3$ strip, dimer correlation decays faster in the bulk than on the boundary.
}
\label{fig:strip_correlation}
\end{figure}

To further characterize the gapless edge mode, we now compute the correlation functions, with the goal to identify scaling dimension of some primary field in the $c=1$ CFT. Here we have to be careful that the conservation of $W_p$ operators put strong constraints on the correlations one can extract~\cite{Baskaran2007,Tan2024}. Indeed, since the local operators $T_i^{x,y,z}$ on site $i$ do not commute with all the $W_p$'s surrounding this site, the standard spin correlation $\langle T_i^{\alpha} T_j^{\beta} \rangle$ would be identically zero beyond nearest neighbor. On the other hand, since $W_p$ operators commute with the Hamiltonian, it is natural to consider the local Hamiltonian term on each bond, i.e., the dimer operator, which fulfills the selection rule imposed by $W_p$. In the following, we consider the dimer operator $T_{i}^x T_{j}^x$ on the $x$-bond, and compute their correlation functions on the strip geometry. The distance in the following is measured along $a_1$ direction, and the result is shown in Fig.~\ref{fig:strip_correlation}.

On an infinitely long strip with width $L_y=2$, we observe that the dimer correlation functions show an exponential decay in the long-distance limit. By fitting the asymptotic behavior using $\mathrm{ln}|C(d)|=-d/\xi_D+\mathrm{const.}$, one can extract the corresponding correlation length. We have checked that this correlation length is the same as the maximal correlation length one would obtain from the leading eigenvalue of transfer matrix of the infinite MPS. And most importantly, $\xi_D$ does not saturate with increasing bond dimension $\chi$, indicating that the dimer correlation would decay algebraically in the infinite $\chi$ limit.

For finite $\chi$, although eventually the dimer correlation would decay exponentially in the long-distance limit, we can fit the correlation using $|C(d)|\sim d^{-\eta}$ in the $d\le \xi_D$ region, and obtain the critical exponent $\eta$, shown in Fig.~\ref{fig:strip_correlation}(a). Then we extrapolate the critical exponent with bond dimension $\chi$ (shown in Fig.~\ref{fig:strip_correlation}(c)) and find the converged exponent is close to $\eta=2$.
Similar behavior is also observed on the boundary of $L_y=3$ strip, shown in Fig.~\ref{fig:strip_correlation}(b) and (c), where the data corresponds to putting the dimer operator on the boundary of the strip.
For the $L_y=4$ strip, the correlation length grows less quickly with $\chi$, and as a result we are not able to perform the finite $\chi$ scaling for the critical exponent here.

Given the critical dimer correlation observed on strips, one may wonder whether they correspond to edge modes. To understand this better, we now compare the dimer correlation function on the boundary versus in the bulk (i.e., the chain in the middle of the strip). For the relatively small width we can compute with MPS ($L_y=3,4$), the dimer correlation on the boundary and in the bulk would eventually decay with the same correlation length in the long distance limit, since they have the same quantum numbers and should have non-zero overlaps. However, their behavior can be quite different in the relatively short-distance regime. Indeed, from Fig.~\ref{fig:strip_correlation}(d), we find that the bulk dimer correlation decays much more quickly than the boundary one, indicating the gapless mode is on the edge.
The existence of edge correlation with exponent $\eta=2$ puts further constraints on the topological order, which we discuss in the next section.

\section{Constraining the topological order}
\label{sec:topo_order}

We now discuss possible topological order based on the numerical results. From the cylinder and strip calculations we have observed that
\begin{enumerate}
    \item All cylinders up to $L_y=5$ have finite correlation lengths, which appear to even decrease as $L_y$ increases, strongly suggesting that the 2D bulk is fully gapped. It then follows from the 1-form symmetry that the bulk must be topologically ordered.
    
    \item On a cylinder, the ground states obtained within each $\Phi_2$ sector are (quasi)-degenerate, indicating that the whole $\Z_4^{(-1)}$  anyon theory is realized in the system. 
    
    \item The $L_y=2,3,4$ strips appear to be critical with $c\approx 1$. We also find that the correlation function of local Hamiltonian terms on the edge decays as $\sim 1/x^2$ where $x$ is the distance along the edge. Both can be naturally accounted for by gapless edge modes on strips described by a $c=1$ chiral CFT with an emergent U(1) symmetry (so the Hamiltonian term flows to the U(1) current operator in the infra-red limit).
\end{enumerate}

Taken together, these results strongly constrain the bulk topological order. Let us now consider the possible anyon theory $\C$~\cite{Kitaev2006, Bonderson2007}. First of all, $\C$ must contain $\Z_4^{(-1)}$ as a subtheory.  Next, the bulk theory must support a $c=1$ chiral CFT with $\U$ symmetry on the edge. Using the classification of $c=1$ rational CFTs, we find that the only possibilities are the $\C=\U_{-2k}$ with $k\in \Z$. For clarity, we will explain the argument below after analyzing the $\U_{-2k}$ theories. 
The anyons in $\U_{-2k}$ are labeled by $[a]$ with $a=0,1,\dots, 2k-1$ mod $2k$, and the $\Z_4^{(-1)}$ theory corresponds to the subgroup generated by $[k/2]$. The exchange statistis of the $[k/2]$ anyon is $e^{-\frac{i\pi (k/2)^2}{2k}}=e^{-i\pi\frac{k}{8}}=-i$, which means $k\equiv 4$ mod 16. Thus the family can be parametrized as $\U_{-8-32m}$ where $m\in \Z$. Compared to the quantum dimension $\cal{D}\approx 3.7$ found from numerically fitting TEE, $\U_{-8}$ is the closet ($\cal{D}=\sqrt{8}\approx 2.8$). We thus postulate $\U_{-8}$ as the ground state topological order.

Let us come back to the reason for singling out $\U_{-2k}$. $c=1$ CFTs consist of three families~\cite{Ginsparg:1987eb}: the ``circle branch" of the $\U_{-2k}$ theories, the ``orbifold branch" which are the orbifolds of the circle branch by the $\Z_2$, and several exceptional cases which are orbifolds of the SU(2)$_1$ theory. Examing the anyon content of orbifold theories, none of them contain $\Z_4^{(-1)}$. For example, the group of Abelian anyons in the $\Z_2$ orbifold of $\U_{-2k}$ is either $\Z_2\times\Z_2$ (all of which are bosons) when $k$ is even or $\Z_8$ when $k$ is odd. But the latter case still belongs to the circle branch. Thus we can exclude the orbifold theories based on these considerations.

We should note that due to the large error bar in the TEE, the identification of $\U_{-8}$ is clearly not conclusive. In fact, just based on TEE one can not rule out $m=-1$, i.e. $\U_{24}$ with $\cal{D}\approx 4.9$ (other values of $m$ are unlikely). On the other hand, $\U_{24}$ would imply that within each sector of a definite $\Phi_2$ value there are six ground states. We have not seen any evidence for such a large ground state degeneracy. Thus we think $\U_{24}$ is less likely, but more work needs to be done to fully settle the issue.

\section{Discussion on generic $\Z_N$ Kitaev models}
\label{sec:ZN_case}

Based on the results for $\Z_3$ and $\Z_4$ Kitaev models, we now make general remarks for all $\Z_N$ cases.

Firstly, the $\Z_N$ 1-form symmetry guarantees that if there is a spectral gap, then the system must have some form of topological order. For odd $N$, the $\Z_N^{(-1)}$ 1-form symmetry already corresponds to a modular anyon theory, so the minimal topological order could just be the $\Z_N^{(-1)}$ theory with $N$ types of anyons. Any topological order realized in this model must contain $\Z_N^{(-1)}$ as a subtheory. Our numerical study of the $\Z_3$ Kitaev model at the ferromagnetic isotropic point identified $\U_{12}$ as the actual topological order~\cite{Chen2024}.

 The situation is quite different for even $N$, because the $\Z_N^{(-1)}$ 1-form symmetry corresponds to a non-modular anyon theory. For $N\equiv 0\pmod{4}$, the element $N/2$ is a transparent boson. Here, transparent means that it braids trivially with all other anyons. A physical topological order realized in a bosonic system does not allow transparent boson other than the identity. There are two options:
 \begin{itemize}
     \item The transparent boson becomes trivial. In this case, the minimal anyon theory compatible with the 1-form symmetry is $\Z_{N/2}^{(-\frac12)}$. 
     \item The $N/2$ boson remains a nontrivial anyon, but then the theory must contain other anyons which braid non-trivially with the $N/2$ boson. The simplest Abelian topological order is $\Z_{2N}^{(-\frac12)}$, which appears to be realized in the $\Z_4$ model.
 \end{itemize}

For $N\equiv 2\pmod{4}$, the transparent element $[N/2]$ is fermionic. In this case, the anyon theory can be factorized into $\Z_2^{(1)}\boxtimes \Z_{N/2}^{(-2)}$. The transparent fermion can not be trivial, so $\Z_2^{(1)}$ must be embedded in a larger anyon theory, as in the original $\Z_2$ Kitaev model.

Secondly, one can notice that for $\Z_N$ Kitaev model ($N\ge 3$), the time-reversal symmetry is broken. This give us a hint that the topological order could be a chiral one. 

Numerically, in both $\Z_3$ and $\Z_4$ Kitaev models we observed a $c=1$ gapless modes on a strip geometry, which are most likely due to chiral edge modes. It is thus instructive to consider possible TOs compatible with the $\Z_N$ 1-form symmetry and with $c=1$ chiral edge modes.
The simplest such theory is $\U_{-2N}$ for both even and odd $N$. 


In fact the $\Z_N$ Kitaev models belong a more general class of models introduced in Ref.~\cite{Ellison2022}. In that case, the models are obtained from ``gauging out" certain anyon types in Pauli topological stabilizer models, which results in a much richer topological phases than the stabilizer code. It would be interesting to explore the phase diagram of that types of models, following the same approach here.

From a technical point of view, this class of models also brings some challenges to tensor network methods. In this study we have solely used MPS based numerical techniques to simulate the two-dimensional model. It would be interesting to apply the projected entangled-pair state to this type of models. In that case, encoding the $\Z_N$ 1-form symmetry into PEPS would be crucial. An important step in this direction was carried out very recently~\cite{Tan2024b}.
It would be interesting to apply this technique to the $\Z_4$ Kitaev model.

To summarize, we have studied a $\Z_4$ generalization of the Kitaev honeycomb model. Combining the $\Z_4$ 1-form symmetry inherent in this model and MPS study on cylinder and strip geometries, we have deduced that this model realizes a U$(1)_{-8}$ chiral topological order at the isotropic ferromagnetic point. For the experimental realization and questions we mentioned above, we hope to continue the study in the near future.

\section*{Acknowledgments}

Part of the MPS calculations were performed using the TeNPy Library (version 0.9.0)~\cite{Hauschild2018}.
This work is supported by  National Natural Science Foundation of China (Grant No.~12304186), Open Research Fund Program of the State Key Laboratory of Low-Dimensional Quantum Physics (Project No.~KF202207), Fundamental Research Funds for the Central Universities, Sun Yat-sen University (project No.~23qnpy60), Innovation Program for Quantum Science and Technology 2021ZD0302100, Guangzhou Basic and Applied Basic Research Foundation (Grant No.~2024A04J4264), and Guangdong Basic and Applied Basic Research Foundation (Grant No.~2024A1515013065).
M.C. acknowledges support from NSF under award number DMR-1846109.
The calculations reported were performed on resources provided by the Guangdong Provincial Key Laboratory of Magnetoelectric Physics and Devices, No.~2022B1212010008.

\appendix

\section{$\Z_N^{(p)}$ anyon theory}
\label{anyontheory}

Here we review the $\Z_N^{(p)}$ anyon theory, following the notations in Ref.~\cite{Bonderson2007}. There are $N$ types of anyons labeled by $[a]$ with $a=0,1,\dots, N-1$ defined mod $N$. Here $[0]$ is the identity anyon, corresponding to local bosonic excitations. The fusion rules between the anyons are given by
\begin{equation}
    [a]\times [b]=[a+b].
\end{equation}

In order for the statistics of anyons to be well-defined, $p$ is required to be an integer for odd $N$, and can be an integer or a half-integer for even $N$. The exchange statistics reads
\begin{equation}
    \theta(a)=e^{\frac{2\pi i p}{N}a^2}.
\end{equation}
The mutual braiding phase between $[a]$ and $[b]$ is $e^{\frac{4\pi ip}{N}ab}$. 

We note that for even $N$, $\Z_N^{(1/2)}$ describes the topological order of the $\U_{N}$ Chern-Simons theory. 

An anyon is said to be transparent if the braiding between an anyon and all other anyons is trivial. An anyon theory is modular if the only transparent anyon is the identity. The anyon theory that arises from a gapped ground state in a bosonic (spin) system must be modular~\cite{Kitaev2006}.
It is easy to see that the $\Z_N^{(p)}$ theory is modular if and only if $\text{gcd}(p,N)=1$ for odd $N$, or $\text{gcd}(2p, N)=1$ for even $N$. In this work we are only interested in the $p=-1$ case. Then for odd $N$ the theory is modular. When $N$ is even, it is readily seen that $[N/2]$ is transparent, with statistics $\theta([N/2])=(-1)^{N/2}$.

\section{single chain limit}
\label{sec:Z4Chain}

In the study of Kitaev material, it has been found that the single chain limit can provide interesting insight into the higher dimensional counterpart~\cite{Wang2020,Sorensen2021}. This is partly due to the fact that many theoretical tools exist in one dimension, e.g., bosonization and conformal field theory technique.
Therefore, here we take a detour to look at the one dimensional limit of this model, namely, a $\Z_4$ Kitaev chain. This could help us to gain some intuition about the microscopic details, and may also be of independent interest.

As shown in Fig.~\ref{fig:Z4KitaevModel}(b), the Hamiltonian for the $\Z_4$ Kitaev chain with $L$ sites is given by:
\begin{equation}
\label{eq:1D}
    H_{1D} = \sum_{j=1}^{L/2} J_x T_{2j-1}^x T_{2j}^x + J_y T_{2j}^y T_{2j+1}^y + \mathrm{h.c.},
\end{equation}
where we have assumed $L$ even. Similar to the two dimensional case, one can notice that this model has an extensive number of conserved quantities, given by
\begin{equation}
    W_{2j}=(T_{2j}^x)^\dag T_{2j+1}^x,\quad W_{2j-1}=(T_{2j-1}^y)^\dag T_{2j}^y.
\end{equation}
To ease the discussion, we further define
\begin{equation}
    V_{2j}=T_{2j}^y T_{2j+1}^y, \quad V_{2j-1}=T_{2j-1}^x T_{2j}^x,
\end{equation}
which are in fact the local terms in the Hamiltonian Eq.~\eqref{eq:1D}.

Using the commutation relations between $\Z_4$ operators, one can find that the $V,W$ operators satisfy the following operator algebra:
\begin{equation}
\begin{split}
    &V_n^4=1,\quad V_{2j}V_{2j\pm 1}=-i V_{2j\pm 1}V_{2j},\\
    &V_m V_n=V_n V_m,\quad \mathrm{for}\ |m-n|>1,\\
    &W_n^4=1,\quad W_{2j}W_{2j\pm 1}=-i W_{2j\pm 1}W_{2j},\\  
    &W_m W_n=W_n W_m,\quad \mathrm{for}\ |m-n|>1,\\
    &V_i W_j=W_j V_i,\quad \forall i,j
\end{split}
\label{eq:algebra}
\end{equation}
Therefore, the subalgebra generated by $V$'s can be represented on a $\Z_4$ spin chain of $L/2$ sites, and the same is true for the subalgebra generated by $W$'s.

Notice that for $L=0$ mod $4$, the following identities holds under periodic boundary condition:
\begin{equation}
\begin{split}
V_2V_4^\dag V_6V_8^\dag\cdots V_L^\dag &= W_1W_3^\dag W_5 W_7^\dag \cdots W_{L-1}^\dag, \\
V_1V_3^\dag V_5 V_7^\dag \cdots V_{L-1}^\dag &=W_2^\dag W_4 W_6^\dag W_8\cdots W_L.
\end{split}
\label{eq:constraint}
\end{equation}
Since each of the product commutes with the Hamiltonian Eq.~\eqref{eq:1D}, therefore the Hamiltonian Eq.~\eqref{eq:1D} has a $\Z_4\times \Z_4$ symmetry. 

The operator algebra in Eq.~\eqref{eq:algebra} suggests that one can map the $V$'s to a dual $\Z_4$ spin chain, whose Pauli operators on each site are denoted as $\sigma$ and $\tau$ with commutation relation $\sigma\tau=i\tau\sigma$~\footnote{The explicit expression for $\sigma$ ($\tau$) is the same as $T^z$ ($T^x$)}, in the following way:
\begin{equation}
    V_{2j}=(\sigma_j^\dag\sigma_{j+1})^{(-1)^j},\ V_{2j-1}=\tau_j^{(-1)^j}.
\label{eq:dualspin}
\end{equation}
This definition works on an infinite chain. Here we have chosen a staggered mapping due to the constraint Eq.~\eqref{eq:constraint}, which one can verify by substitution.
For PBC, for the mapping to be well-defined we should have $L\equiv 0\ \mathrm{mod}\ 4$. In addition, the constraint Eq.~\eqref{eq:constraint} means that the $\sigma$ spins should be allowed to have twisted boundary conditions. Thus more generally we should replace Eq.~\eqref{eq:dualspin} with the following mapping:
\begin{equation}
\begin{split}    
    & V_{2j}=(\sigma_j^\dag\sigma_{j+1})^{(-1)^j},\ 1\le j < L/2\\
    & V_L=\nu \sigma_{L/2}^\dag\sigma_1,\\
    & V_{2j-1}=\tau_j^{(-1)^j}
\end{split}
\label{eq:map_V}
\end{equation}

Similarly, for the subalgebra generated by $W$'s, we can have another mapping:
\begin{equation}
\begin{split}
    & W_{2j}=(\tilde{\sigma}_j^\dag\tilde{\sigma}_{j+1})^{(-1)^j},\ 1\le j <L/2\\
    & W_L = \tilde{\nu}\tilde{\sigma}_{L/2}^\dag\tilde{\sigma}_1,\\
    & W_{2j-1}=\tilde{\tau}_j^{(-1)^j}.
\end{split}
\label{eq:map_W}
\end{equation}

With the mappings Eqs.~\eqref{eq:map_V} and \eqref{eq:map_W}, the constraints Eq.~\eqref{eq:constraint} become
\begin{equation}
    \nu=\prod_j\tilde{\tau}_j,\ 
    \tilde{\nu}=\prod_j\tau_j^\dag.
\end{equation}

When $L\equiv 0$ mod $4$, the dual Hamiltonian takes the form of the standard $\Z_4$ clock model:
\begin{equation}
\begin{split}
    H_{\rm dual}=&\sum_{j=1}^{L/2-1}(J_y\sigma_j^\dag \sigma_{j+1}+J_x\tau_j+\mathrm{h.c.})\\
    &+ (\nu J_y \sigma_{L/2}^\dag\sigma_1+J_x\tau_{L/2}+\mathrm{h.c.}).
\end{split}
\label{eq:dual}
\end{equation}
For $L\equiv 2$ mod $4$, it is mapped to the clock model with a charge-conjugation defect:
\begin{equation}
\begin{split}
    H_{\rm dual}=&\sum_{j=1}^{L/2-1}(J_y\sigma_j^\dag \sigma_{j+1}+J_x\tau_j+\mathrm{h.c.})\\
    &+ (\nu J_y \sigma_{L/2}\sigma_1+J_x\tau_{L/2}+\mathrm{h.c.}).
\end{split}
\label{eq:defect}
\end{equation}
This implies that the translation ($j\rightarrow j+1$ on the Kitaev chain) acts on the fields nontrivially as charge conjugation. This point is missing in the single chain analysis in Ref.~\cite{Barkeshli2015}.

The $\Z_4$ clock chain has been studied extensively in the literature, and the phase diagram is well-known for $J_x,J_y\in \mathbb{R}$. Most importantly, when $J_x=J_y$, the single chain is critical and described by a central charge $c=1$ Ising$^2$ conformal field theory.

Above analysis in fact also works for the $\Z_3$ case, where an analysis for the single chain was carried out in Ref.~\cite{Barkeshli2015} using parafermion operators. Here our approach does not involve parafermion, and the mapping is made transparent in the spin language.

\bibliography{bibliography}

\end{document}